\documentclass{svproc}
\usepackage{url}
\usepackage{graphicx}

\begin{document}
\mainmatter              
\title{Study and security analysis of the Spanish identity card}
\titlerunning{Study and security analysis of the Spanish identity card}  
\author{Javier Correa-Marichal\inst{1} \and Pino Caballero-Gil\inst{1}
\and Carlos Rosa-Remedios\inst{2} \and Rames Sarwat-Shaker\inst{3}}
\authorrunning{Javier Correa-Marichal et al.} 
\tocauthor{Javier Correa-Marichal, Pino Caballero-Gil, Carlos Rosa-Remedios, and Rames Sarwat-Shaker}
\institute{ Department of Computer Engineering and Systems, University of La Laguna,  38200 Tenerife, Spain\\
\email{alu0101233598@ull.edu.es},\email{pcaballe@ull.edu.es},
\and
CECOES 1-1-2, Canary Islands Government,
\\
38003 Tenerife, Spain\\
\email{crosa@gsccanarias.com},
\and
Telefónica Tech,\\
28050 Madrid, Spain\\
\email{rames.sarwatshaker@telefonica.com}}

\maketitle              

\begin{abstract}
The National Identity Document is a fundamental piece of documentation for the identification of  citizens throughout the world. That is precisely the case of the  DNI (Documento Nacional de Identidad) of Spain. Its importance has been enhanced in recent years with the addition of a chip for the authentication of users within telematic administrative services. Thus, the document has since been called: electronic DNI or simply DNIe. Sensitive user information is stored in that integrated circuit, such as personal and biometric data, along with signature and authentication certificates. Some of the functionalities of the DNIe in its current version at the time of writing this work have been implemented for years in the DNI 3.0 version launched in 2015, and therefore have already been extensively studied. This work provides a theoretical and practical compilation study of some of the security mechanisms included in the current DNIe and in some of the applications that require its use. It has been carried out using only mobile devices and generic card readers, without having any type of privileged access to hardware, software or specific documentation for the interception of packets between the DNIe and the destination application. In other words, it is an exploratory analysis carried out with the intention of confirming with basic tools the level of robustness of this very important security token.
\end{abstract}
\section{Introduction}
The Spanish identity card or DNI (Documento Nacional de Identidad) is a document issued by the Spanish Ministry of the Interior, to prove the identity and personal data of any Spanish citizen. Since obtaining it is mandatory for all Spaniards over 14 years of age, millions of DNIs are issued annually in Spain \cite{anuario}. Taking advantage of its popularity, and in order to promote the digitization of telematic services offered by public administrations, in 2006 a new version of the DNI was launched  with a chip that offered various functionalities related to the identity of the holder, and since then it has been called electronic DNI or DNIe.

The security of the physical document and its electronic components and related software is improved with each new revision.
Each update of the DNIe, before being certified by the Certification Body of the National Cryptologic Center, passes an evaluation process developed by the National Currency and Stamp Factory – Royal Mint under request of the General Police Directorate, and carried out by an accredited laboratory that passes SOG-IS audits. This evaluation  follows the Common Criteria methodology (ISO/IEC 15408).
Specifically, the DNIe software has been certified with the evaluation assurance level EAL4+ EAL4 AVA\_VAN.5, and the chips have been certified as a Secure Signature Creation Device, in accordance with European standards \cite{BOE}. These certifications provide a very high level of confidence. However, design or implementation errors often go unnoticed in certified products already deployed on technologies, so it is always necessary to consider security as a process and not as a state.

In the launch of the DNIe 3.0 in 2015, an interface for use through NFC was included, which allows the DNIe to be used directly through mobile devices that include this technology, in an effort to popularize its use \cite{dnimuerte}.

At the time of writing this work, the last revision of the DNI had been launched in August of the previous year, in 2021, with an announcement on the official website \cite{nuevodni}. One of the most notable features of this version is the design and functionality, seeking to homogenize the identity documents of the countries of the European Union so that their use can be standardized and approved according to the eIDAS regulation, of digital identification in Europe. In addition, as announced on the police website  \cite{policia}, the current version of the DNIe includes new security measures, both visible and invisible.

This study has been performed on the latest version of the DNIe since it was carried out on a document issued at the end of 2021. It is  an exploration of the implementation of  the functionalities  and   security mechanisms contained in the document and some linked apps. 
The main goal of this research has been to confirm the resilience of the DNIe against several frequent    attack vectors  on electronic identity cards and NFC devices \cite{Arroyo}.
In fact, this study does not provide new vulnerabilities or attacks on the contactless protocols of the DNIe, but rather   serves to show that, in general, the protocols used to communicate through the contactless interface with the DNIe are sufficiently secure.
Following the rules of ethical hacking, this analysis  has been presented to the Spanish National agency in charge of developing the DNIe.

This document is organized as follows. Sections II, III and IV introduce, respectively, the different interfaces of the DNIe,  the logical structure of the data stored in the DNIe and the security mechanisms developed to protect its integrity. Section V collects some details of the research carried out on the current version of the DNIe and various applications that require its use. Finally, section VI presents the conclusions of this work.
\section{User interface}
The DNIe chip is accessible through a dual interface that supports access through a standardized contact point such as ISO/IEC 7816 \cite{iso78162} and wirelessly through an NFC antenna, following the set of protocols defined in the ISO/IEC 14443 standard \cite{iso144433}.
To carry out the present study, the NFC interface has been chosen because it is compatible with Android devices with this technology. The ISO/IEC 14443 standard is in turn supported by the application layer level protocol defined in ISO/IEC 7816-4, where communication is carried out through pairs of commands and responses called {\it Application Protocol Data Units} (APDUs). In this standard, a file system and the commands needed to perform queries are defined. Several applications can be found on the chip, separated in different \textit{Dedicated Files} (DF) that hang from the root of the file system, signaled by the \textit{Master File} (MF). In these, a collection of \textit{Elementary Files} (EF) is stored, where the chip data is saved.

For European identity documents with an optional electronic passport application, as is the case of the DNIe, it is specified in the BSI TR-03110-4 standard that, in order to ensure interoperability as an eIDAS token,  applications for \textit{eID} and  \textit{eSign} must be implemented; and, optionally, also the \textit{ePassport} app \cite{tr031104}.

\section{Logical data structure} 
The ICAO 9303 standard establishes the different aspects of {\it electronic Machine Readable Travel Documents} (eMRTD). To ensure the international interoperability of the data stored in the DNIe, this standard defines a Logical Data Structure (LDS) for all electronic passports. The stored information is assigned during the process of creating the DNIe and cannot be modified later, serving as a protection mechanism for the holder's personal information against possible manipulation attacks. These data are accessible through the application \textit{ePassport} of the DNIe.

The data stored in the DNIe includes personal information about the holder such as biometric data, together with elements used during the execution of the security protocols established to guarantee the legitimacy of the document and the stored information.

There are two other mandatory files that contain information about the logical data structure. The \textit{EF.COM} file contains the versions of LDS and the Unicode character encoding standard used, as well as a list of the data groups present in the application. On the other hand, the file \textit{EF.SOD} contains, for each of the data groups included in the DNIe, their hashes and digital signatures made by the \textit{Document Signer} (DS) of the DNIe \cite{icao930310}.

\section{Security mechanisms}

Part 11 of the ICAO 9303 standard defines a set of security measures to increase the resilience of electronic passports against the most common attacks against this type of document \cite{icao930311}. Table \ref{tab:sectable} includes a summary of these security protocols, indicating the technique used in each one to mitigate each specific attack. The protocols that are of greatest interest in this research, as they are implemented in the current DNIe, are described  below.

\begin{table*}[htb]
\centering
\caption{Security algorithms implemented in eMRTD documents}
\label{tab:sectable}
\begin{tabular}{lll}
\hline
\hline
Protocol  & Technique & Attack \\ \hline
Basic Access Control (BAC) & Authentication and  & Information Theft \\
 & secure channel &  \\
Supplemental Access Control (SAC)  & Authentication and  & Information theft \\
 & secure channel &  \\
Passive Authentication (PA) & Digital signature & Forgery,  \\
 &  & data manipulation \\
Active Authentication (AA) & Challenge-response & Cloning \\
Chip Authentication (CA) & Authentication & Cloning \\
Terminal Authentication (TA) & Authentication via PKI & Information theft \\
\hline
\hline
\end{tabular}
\end{table*}

\subsection{Basic/Supplemental Access Control}

The purpose of the \textit{Basic Access Control} (BAC) protocol is to guarantee that the \textit{Inspection System} (IS) has maintained visual contact with the document, thus protecting the data of a sensitive nature stored in the DNIe. This security mechanism has a dual purpose. First, it makes it more difficult to steal information through the card's contactless interface without the cardholder's knowledge. Second, it establishes a secure channel through which all subsequent traffic can be sent, protecting the communication from eavesdropping attacks and data corruption.
For the execution of this protocol, it is first necessary for the IS to obtain the \textit{Document Basic Access Keys} that are derived from the \textit{Machine Readable Zone} (MRZ) of the DNIe. In particular, three fields are obtained from it.

In order to correct weaknesses reported in the BAC protocol \cite{dniassess}\cite{epassportsurvey}, the use of \textit{Supplemental Access Control} (SAC) is proposed as a substitute. This authentication mechanism uses the \textit{Password Authenticated Connection Establishment} (PACE) protocol to provide strong encryption and allow the sharing of \textit{Message Authentication Codes} (MAC) for secure channel establishment.

\subsection{Passive Authentication}

\textit{Passive Authentication} (PA) is a security mechanism that allows verifying the authenticity and integrity of the information stored in the LDS. During the personalization phase of the chip in the process of issuing the DNIe, the hashes corresponding to the different groups of data stored in the LDS are calculated and stored in the file \textit{EF.SOD}, together with a digital signature made on them. In the document inspection process, it is a task of the IS  to consult the \textit{EF.SOD} file and validate the hashes for each of these groups, and verify that the certificate of the \textit{Document Signer} used is signed, in turn, by a recognized and valid \textit{Country Signing Certification Authority}.

\subsection{Extended Access Control}

The handling of biometric data stored in the DNIe, such as the holder's fingerprint information, is considered more sensitive than that of other data stored in the document. That is why there is a more restrictive procedure for its access. This protocol, called \textit{Extended Access Control} (EAC), is specified in the BSI document TR-03110 for European passports and identity cards \cite{tr031102}. It consists of three parts: passive authentication,  \textit{Chip Authentication} (CA), and  \textit{Terminal Authentication } (TA).

\section{Security Assessment}
The research carried out on several apps that require the use of the DNIe is introduced below.

\subsection{ePassport}

In the \textit{ePassport} application, personal and biometric data of the holder of the DNIe are stored, structured within the aforementioned LDS, standardized for  eMRTD documents. Illegitimate access by an attacker to this functionality, through the RFID capability enabled in the DNIe, could reveal personal information about an individual without their consent, allowing potential access to highly sensitive data such as the user's biometric data.

In order to prevent arbitrary reading of sensitive data stored in the DNIe, security protocols such as BAC, PACE or EAC, detailed in the previous section, are implemented in the card. Due to their importance within the security framework of the DNIe, an exploratory study has been carried out
to confirm the robustness of its implementation in the document.
For this, the Android application development kit provided by the National Police  \cite{codigofuente} has been used, based on the implementation of the \textit{DNIeDroid} library. Through the use of the \textit{Frida}  tool \cite{frida}, it would theoretically be possible to dynamically modify the behavior of the methods defined in that library and replace them entirely with user-coded functions, through a technique called \textit{API Hooking}. Taking advantage of this functionality, it is possible to hook up the methods responsible for establishing secure communication with the DNIe, allowing the APDU packets sent between the terminal and the document to be analyzed and even modified in real time.

Through the malformation of packets sent between both devices, the possibility of finding failures in the implementation of the applications that make up the DNIe has been studied. Specifically, in the experiments, special attention has been paid to the implementation of the PACE protocol, studying the robustness of the most critical operations carried out in it, such as the generation of cryptographically random numbers  \cite{dniassess}, or the final validation of the ephemeral key of the negotiated session.

In another line of research, the feasibility of exploiting known vulnerabilities that affect NFC devices has been analyzed.
In this sense, the experiments have shown that most of the attacks described in the literature, such as eavesdropping or data manipulation, are mitigated in the DNIe through the implementation of the security measures described in the previous section. 

\begin{figure}
\centering \includegraphics[width=10
cm]{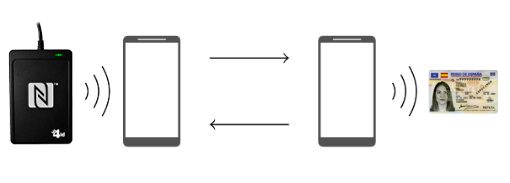} 
\caption{Graphical representation of a \textit{relay} attack}
\label{fig:relay}
\end{figure}

The countermeasures implemented for  attacks known  as \textit{relay} attacks on NFC devices, have been specifically analyzed in this work.
\textit{Relay} attacks, described in \cite{relayattack} as “mafia fraud” and in \cite{wormholeattack} as “wormhole attack”, follow the \textit{Man- in-the-Middle} technique and consist of the extension of the communication between two participants, using two rooted devices to forward the command and response APDU packets between the card and the reader. Since these packets can be sent over the Internet, both devices could be far apart, thus violating the maximum theoretical distance restriction for reading NFC cards (see Figure \ref{fig:relay}). Given that encrypted messages are involved in \textit{relay attacks}, some cryptographic methods could be used to detect them. An example of a solution proposed in the literature is through distance-bounding authentication protocols \cite{DBP}.

The conclusion from the experiments on the \textit{ePassport} app is that the security measures discussed in the previous section help prevent eavesdropping and data corruption attacks. Regarding \textit{relay} attacks, using the \textit{NFCGate} \cite{nfcgate} app for Android devices, it is possible to perform a \textit{relay} attack between a terminal with a reader application built on the \textit{DNIeDroid} library and a DNIe.

\subsection{eSign}

The access to the \textit{eSign} application is similar to that of \textit{ePassport}, since it is necessary to carry out an execution of the PACE protocol using the PIN associated with the DNIe as a shared secret between the card and the reader. Once this protocol has been successfully executed, a secure channel is established and access to the signing functionality is allowed. The operations that use these certificates are always executed within the DNIe, ensuring that the private keys never leave the memory of the document, so that the DNIe can be considered a secure signature creation device.

Given the few existing attack vectors against this functionality, for this research it has been decided to analyze the official software implementation for the interaction of a computer with a document using the  digital signature tool called  @firma  suite \cite{clienteafirma}. @Firma is an electronic signature client application that runs on the user's PC.
As is traditional in cryptology, in the study carried out it was decided to consider theoretical, and usually non-practical, hypothetical conditions, in the style of attacks with known plaintext. Specifically, it is considered the hypothesis  that an attacker is capable of extracting the PIN from the DNIe through the interaction of a legitimate user with the @firma application. Under these conditions, it was possible to change the document to be signed dynamically during the execution of the application, forcing the user to legally commit to contracts chosen by the attacker, through malware running on the user's PC, which confirms the need to protect the security of used PCs.

In the research carried out, it was decided to explore the viability of attacks by modifying the @firma client as little as possible. As future work, changes could be considered in the source code of this application \cite{afirmagithub} to add a malicious routine that would take care of replacing the document to be signed by the user without their consent. However, this approach would require administrative permissions on the computer to be able to replace the application binary, thus implying greater complexity in exploiting the attack.

The approach followed in this case study has been based primarily on the use of the Java JDB debugger included in the \textit{Java Development Kit} tools. With this debugger it is possible to connect to a running process in order to use breakpoints located in strategic functions to observe or modify the local variables stored in memory, thus being able to use a PIN entered by the user, for the subsequent initiation of a non-signature. authorized. It would also be possible to dynamically modify the file to be signed using the same method, although in this case the tools provided in the JDB environment are somewhat limited in handling complex data structures.

In this research, a proof of concept of a script capable of recovering the PIN entered by a user in the @firma application has been developed, without the need to modify the original binary within a Linux environment. In addition, the feasibility of an arbitrary document signature attack using the described modifications has been studied.

\section{Conclusions}

In this paper, an exploratory study of the security implemented in the DNIe in its current version has been presented, as well as of some  applications that require its use.
The line of research  has been eminently practical, in order to confirm the robustness of the technology and software that surround the DNIe.

In general, the high level of security of the DNIe has been confirmed. However, the introduction of NFC technology in the DNIe from version 3.0, although it offers a better user experience, exposes the DNIe to possible  attacks such as \textit{relay}, which although they cannot be used to access sensitive data of the DNIe, they do allow the basic data of the DNIe to be read from an indeterminate distance. On the other hand, among the applications that require the use of the DNIe, the @firma client could become a weak point within the chain of trust if the impossibility of capturing the user's PIN is not totally  guaranteed. In our view, both issues could be potential threats if certain types of malware gain access to the document.
Thus, in conclusion, it can be stated that in general today the DNIe can be considered a secure platform with no active threats against its security. However, it is always necessary to take into account the great impact that the possible creation and distribution of specific malware would have, taking advantage of the weaknesses of the software to commit the victims in legal settlements without their consent. Therefore, although the hardware and software infrastructure of the DNIe has proven to be sufficiently solid against the most direct attacks, it is important to favor the permanent audit of the security of all the related applications, since they could become a key entry point for attackers who seek to misuse the functionalities of the DNIe.

\section*{Acknowledgments}

This research has been supported by the Spanish Ministry of Science, Innovation and Universities, the State Research Agency and the European Regional Development Fund under project RTI2018-097263-B-I00. The authors also appreciate the feedback received from the Fábrica Nacional de Moneda y Timbre-Real Casa de la Moneda.
%
%

\end{document}